%% file: main.tex
\documentclass[conference,a4paper]{IEEEtran}
\usepackage{xcolor}
\usepackage{balance}
\usepackage{siunitx}

\definecolor{rose_color}{RGB}{219, 48, 122}
\definecolor{blue_color}{RGB}{72, 159, 240}
\definecolor{green_color}{RGB}{56, 232, 79}
\definecolor{grey_color}{RGB}{50, 76, 80}  %

\colorlet{los_color}{rose_color}
\colorlet{rup_color}{blue_color}
\colorlet{rdw_color}{green_color}
\colorlet{tx_color}{grey_color}
\ifCLASSINFOpdf
  \usepackage[pdftex]{graphicx}
\else
  \usepackage[dvips]{graphicx}
\fi

\usepackage{tikz}
\usetikzlibrary{intersections}
\tikzstyle{block}=[draw=grey_color, very thick, align=center, rounded corners, minimum height=1.2cm, fill=blue_color, text=black]

\usepackage{pgfplots, pgfplotstable}
\pgfplotsset{compat=1.18}
\usepgfplotslibrary{groupplots}
\usepgfplotslibrary{statistics}

\usepackage{amsmath,amssymb}

\ifCLASSOPTIONcompsoc
 \usepackage[caption=false,font=normalsize,labelfont=sf,textfont=sf]{subfig}
\else
 \usepackage[caption=false,font=footnotesize]{subfig}
\fi

\usepackage{url}
\usepackage{hyperref}

\usepackage{glossaries-extra}

\setabbreviationstyle[acronym]{long-short}

\glssetcategoryattribute{acronym}{nohyperfirst}{true}

\newacronym{v2v}{V2V}{Vehicle-to-Vehicle}
\newacronym{rt}{RT}{Ray Tracing}
\newacronym{dynrt}{DynRT}{Dynamic \gls{rt}}
\newacronym{diffrt}{DiffRT}{Differentiable \gls{rt}}
\newacronym{go}{GO}{Geometrical Optics}
\newacronym{utd}{UTD}{Uniform Theory of Diffraction}
\newacronym{los}{LOS}{line of sight}
\newacronym{em}{EM}{electromagnetic}
\newacronym{mlm}{MLM}{Multipath Lifetime Map}
\newacronym{tx}{TX}{transmitter}
\newacronym{rx}{RX}{receiver}
\newacronym{ad}{AD}{automatic differentiation}
\newacronym{ml}{ML}{Machine Learning}
\newacronym{sixb}{6B}{6-Building}
\newacronym{twob}{2B}{2-Building}
\newacronym{sha256}{SHA256}{256-bit Secure Hash Algorithm}

\usepackage{fancyhdr}
\begin{document}

\title{Comparing Differentiable and Dynamic Ray Tracing: Introducing the Multipath Lifetime Map}

\author{\IEEEauthorblockN{
Jérome Eertmans\IEEEauthorrefmark{1},
Enrico Maria Vitucci\IEEEauthorrefmark{2},
Vittorio Degli-Esposti\IEEEauthorrefmark{2},
Laurent Jacques\IEEEauthorrefmark{1},
Claude Oestges\IEEEauthorrefmark{1}
}                                     %
\IEEEauthorblockA{\IEEEauthorrefmark{1}%
ICTEAM, Université catholique de Louvain, Louvain-la-Neuve, Belgium, \href{mailto:jerome.eertmans@uclouvain.be}{jerome.eertmans@uclouvain.be}}
\IEEEauthorblockA{\IEEEauthorrefmark{2}%
DEI, University of Bologna, Bologna, Italy}
}

\maketitle

\thispagestyle{fancy}

\begin{abstract}
With the increasing presence of dynamic scenarios, such as Vehicle-to-Vehicle communications, radio propagation modeling tools must adapt to the rapidly changing nature of the radio channel. Recently, both Differentiable and Dynamic Ray Tracing frameworks have emerged to address these challenges. However, there is often confusion about how these approaches differ and which one should be used in specific contexts. In this paper, we provide an overview of these two techniques and a comparative analysis against two state-of-the-art tools: 3DSCAT from UniBo and Sionna from NVIDIA. To provide a more precise characterization of the scope of these methods, we introduce a novel simulation-based metric, the Multipath Lifetime Map, which enables the evaluation of spatial and temporal coherence in radio channels only based on the geometrical description of the environment. Finally, our metrics are evaluated on a classic urban street canyon scenario, yielding similar results to those obtained from measurement campaigns.
\end{abstract} %

\vskip0.5\baselineskip

\section{Introduction}

As telecommunication systems operate in increasingly dynamic environments, such as those found in \gls{v2v} communications, the need for radio propagation models capable of handling rapidly varying channel conditions has become essential. These dynamic scenarios introduce significant challenges for propagation models, which must not only maintain high accuracy but also adapt efficiently to continuously changing environments.

Historically, accurate modeling techniques like \gls{rt} and the Method of Moments are static, requiring recalculations for even slight changes in scene configuration, such as the displacement of a \gls{rx}. To address the challenges imposed by dynamic scenarios, \gls{dynrt} was introduced as a way to extrapolate future time instants of a moving scene based on a static Ray Tracing simulation and manually derived derivatives. More recently, automatic differentiation techniques have enabled the development of \gls{diffrt} software. Although \gls{dynrt} and \gls{diffrt} share some conceptual similarities, they differ significantly in design and objectives, which often leads to confusion, especially as \textbf{few papers directly compare the two approaches}. In regard to the scope for which dynamic scene extrapolations remain valid, this is primarily estimated from measurements, e.g., with the stationarity analysis in \gls{v2v} scenarios, but this \textbf{does not extend well to complex changes in the scene or in the radio materials}.

In alignment with the aforementioned challenges, we summarize our major contributions as follows.
\begin{enumerate}
    \item We provide a \textbf{first qualitative comparison of two state-of-the-art tools}, 3DSCAT \cite{threedscat}, a \gls{rt} tool developed at UniBo that incorporates \gls{dynrt} \cite{drt-2020}, and Sionna \cite{sionnart}, an open-source simulation framework from NVIDIA that features a \gls{diffrt} engine for radio propagation modeling; 
    \item We introduce a \textbf{novel simulation-based visual tool}, the \gls{mlm}, along with two metrics, to accurately determine when \gls{dynrt} can be applied. In contrast to decisions based on measurements, \gls{mlm} provides a solution that is independent of frequency and accounts only for the geometry of the scene. This makes it possible to extend the solution to any radio-material properties.
\end{enumerate}

The remainder of this paper is organized as follows. First, we review the key components of modern ray tracing techniques and the fundamental principles underlying them. We define the terms \gls{dynrt} and \gls{diffrt} as they are used in this work and provide relevant historical context. Next, we conduct a comparative analysis of two state-of-the-art \gls{rt} frameworks, 3DSCAT and Sionna, evaluating qualitative aspects based on their respective implementation. We then introduce our novel visual comparison tool, \gls{mlm}, along with its associated metrics, and demonstrate its application in an urban street canyon scenario. Finally, we summarize the results presented and suggest future directions for research.

\section{Ray Tracing for Radio Propagation}

In the section, we briefly recall the important concepts behind \gls{rt} in the context of radio propagation. Then, we review the two methods of interest, \gls{dynrt} and \gls{diffrt}, by layering the fundamental principles.

\subsection{Fundamentals}

From a virtual representation of a given environment, known as a scene, a \gls{rt} tool simulates wave propagation by dividing an emitted wavefront into a set of rays. These rays interact with objects in the scene, and by applying principles from \gls{go} and the \gls{utd} \cite{threedscat}, the exact coordinates of the ray paths can be determined. Depending on the desired balance between accuracy and computational efficiency, there are two primary families of methods for solving ray paths.

For higher accuracy, exhaustive \gls{rt} methods attempt to enumerate all possible ray paths using exact techniques, such as the image method for specular reflection or minimization-based methods for more complex interactions like diffraction \cite{threedscat, fpt, mpt}. On the other hand, ray-launching offers greater computational efficiency by emitting multiple rays from a specified \gls{tx} node \cite{threedscat}. Each ray is allowed a fixed number of interactions with the scene's objects before being terminated. The received signal is then approximated from the ray paths that fall within the vicinity of each \gls{rx}.

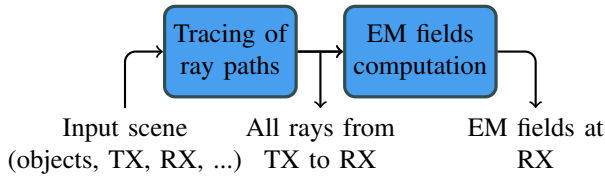
\begin{figure}
  \centering
  \vspace{+0.3mm}
  \input{tikz/pipeline}
  \caption{Simplified two-step procedure taken by most \gls{rt} software. First, the rays are traced based on \gls{go} and \gls{utd}. Then, the radio materials and \gls{em} properties are used to compute the \gls{em} fields.}
  \label{fig:pipeline}
\end{figure}

For both approaches, the \gls{em} fields are typically computed in a separate step (see \autoref{fig:pipeline}), as the ray path directions are independent of the frequency or antenna characteristics. A full recalculation of all ray paths for a given scene is referred to as a \textbf{snapshot}.

In dynamic environments, where ray path coordinates may vary smoothly with respect to changes in object positions, there is potential to bypass the need for full snapshot recalculations. Instead, local derivatives could be used to extrapolate the changes in the perceived \gls{em} fields. This concept will be explored further in the following sections.

\subsection{Dynamic and Differentiable Ray Tracing}

Although the first occurrence of the term ``\emph{dynamic ray tracing}'' seems to date back to the early 1980s \cite{drt-seismic}, when it was used to model the propagation of seismic waves, the use of \gls{dynrt} as we define it in this article appeared only in the early 2020s \cite{drt-2020,drt-2021}. Meanwhile, the term has also been borrowed by both the radio propagation \cite{drt-1990} and computer graphics \cite{drt-2000} communities, but for very different purposes. Finally, while \gls{rt} used to study earthquakes involved solving differential equations, \gls{rt} presented today is derived from \gls{go} and \gls{utd} and accurately models wave propagation.

The primary motivation behind \gls{dynrt} comes from a simple observation: in dynamic scenarios, such as when some \glspl{rx} are moving down a street, most of the environment remains static. Traditional \gls{rt} tools do not take this into account, requiring a complete recalculation of a new snapshot every time the scene changes. However, \gls{rt} is structured so that the types of ray paths affecting a particular region of the scene remain consistent within a local neighborhood. The size and shape of these regions are determined by the shadowing and visibility effects of objects in the environment. To address this, \cite{drt-2020} proposed a method for extrapolating new \gls{rt} simulations from a static snapshot as long as the user (or any dynamic object) remains within the same visibility region, i.e., the region where multipath structure remains the same.

As \gls{rt} and \gls{em} field calculations can be treated as separate processes, \gls{dynrt} often reduces to manually deriving the first---and sometimes second-order---derivatives of the ray path coordinates with respect to the variables of interest, such as the position of a moving \gls{rx}. These derivatives can then be used in a Taylor expansion to predict future changes in the ray paths.

Although the concept of \gls{ad} is not new \cite{ad-1980}, the recent rise of \gls{diffrt} software has been propelled by the development of highly efficient, GPU-optimized \gls{ml} libraries such as PyTorch and TensorFlow for Python. At its core, \gls{ad} applies the chain rule and known derivatives to recursively compute exact gradients. These libraries provide a differentiable framework, allowing users to compute the gradient of \textbf{any output} with respect to \textbf{any input} parameter at a computational cost typically proportional to the function being evaluated. This makes them highly effective for \gls{ml} tasks, but they have also gained attention in the radio propagation community, most notably with the release of Sionna \cite{sionnart}.

While \gls{diffrt} could theoretically be used to compute local derivatives for extrapolating future snapshots, similar to \gls{dynrt}, its primary applications lie in optimizing networks through gradient-based methods, and learning material properties of the radio environment from simulation data \cite{radioenvlearning}.

\subsection{Determining the Scope of Snapshot Extrapolation}

As previously stated, the duration over which a snapshot can be extrapolated on the basis of a previous snapshot is closely linked to the concepts of coherence time, denoted as \(T_c\), and stationarity analysis \cite{drt-2020}. However, these quantities are typically obtained through measurements, which makes the process time-consuming, costly, and not particularly scalable to large scenes. Moreover, the results are dependent on the frequency utilized, the type of antenna, and other variables that may vary during a simulation, making it challenging to identify the scope of application of these extrapolations. In \autoref{sec:mlm}, we present a novel approach for determining the scope of application of these extrapolations based solely on a geometry-based \gls{rt} simulation.

\section{Comparison}

In this section, we present a qualitative comparison between \gls{dynrt} and \gls{diffrt}, using 3DSCAT and Sionna as representative software for each method. Since both tools theoretically produce identical derivatives, a quantitative comparison would be less informative. Furthermore, performance differences are more likely influenced by the target hardware or compiler optimizations rather than the methods themselves.

\subsection{Dynamic Ray Tracing with 3DSCAT}

The 3DSCAT software, developed at UniBo, is a \gls{rt} tool written in C++ that outputs ray path coordinates based on a specified scene configuration. Snapshot extrapolation using \gls{dynrt} and subsequent \gls{em} field calculations are performed separately with MATLAB code.

\subsection{Differentiable Ray Tracing with Sionna}

Sionna is a Python-based library for link-level simulations developed by NVIDIA, built on top of TensorFlow, an efficient, GPU-compatible \gls{ad} framework. This framework enables Sionna to compute derivatives for nearly any scene parameter, making it highly suitable for various radio-focused \gls{ml} applications.

\subsection{Pros and Cons}

Differentiating by hand, as typically done with \gls{dynrt}, is laborious and prone to human error. Additionally, derivatives are often obtained under specific assumptions, e.g., a user moving in a fixed area, and thus remain valid only within a local region, limiting their applicability in general optimization contexts.

Conversely, \gls{ad} is appealing because it automates the complex and error-prone process of computing derivatives and often includes built-in solutions compatible with GPUs or TPUs. This allows derivatives to be easily recomputed after significant environmental changes and at various points throughout the scene. However, using \gls{ad} often restricts developers to a limited set of operations, sometimes leading to more intricate programming solutions and requiring the same \gls{ad} framework throughout the codebase. While this consistency is not an issue when developing a new tool, it does mean that existing commercial \gls{rt} software cannot be made differentiable with \gls{ad}. In contrast, \gls{dynrt} allows differentiation to be implemented as a separate module, as demonstrated by 3DSCAT's \gls{dynrt} plugin.

Finally, \gls{diffrt} lacks the interpretability provided by symbolic derivatives, which \gls{dynrt} offers by using explicit symbolic differentiation.

\section{Multipath Lifetime Map}\label{sec:mlm}

\begin{figure*}
  \centering
  \input{tikz/multipath_cells}
  \subfloat[\Glsxtrlong{los}.]{\show_multipath_cells{1}\label{fig:multipath_cells_1}}
  \quad
  \subfloat[Upper wall reflection.]{\show_multipath_cells{2}\label{fig:multipath_cells_2}}
  \quad
  \subfloat[Lower wall reflection.]{\show_multipath_cells{3}\label{fig:multipath_cells_3}}
  \quad
  \subfloat[All types combined.]{\show_multipath_cells{0}\label{fig:multipath_cells_all}}
  \quad
  \caption{Example of multipath cells for individual---and combined---types of interaction, for a fixed \gls{tx} position (circle node) and a moving \gls{rx}. Each color indicates a unique multipath cell, i.e., regions reached by the same type of interaction. When all types are considered simultaneously, more cells are created, each having a different multipath structure. Blank regions are part of the no-multipath cell. \protect\autoref{fig:multipath_cells_all} exhibits 7 unique cells, split across 11 continuous regions.}
  \label{fig:multipath_cells}
\end{figure*}
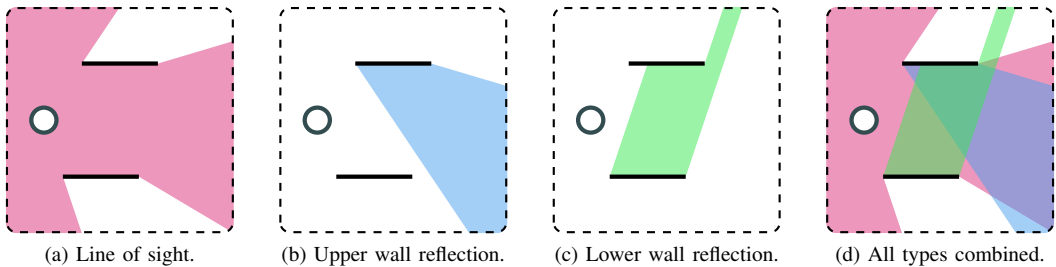

Although it is relatively straightforward to assume that users, or any moving object, will experience the same multipath structure in their immediate vicinity, determining appropriate values for \(T_c\) and understanding how to compute it for a specific scene are less clear. Additionally, using a time-based reference instead of distance implies that the solution is tailored to a specific ``displacement speed''. To address these uncertainties, we introduce a new simulation-based visual tool, named \gls{mlm}, along with associated metrics to offer a deeper understanding of when \gls{dynrt} is most effective.

In the following sections, we define the concepts of similar multipath structures, cells, and the relevant metrics for evaluating them. While we exemplify \gls{mlm} on moving \glspl{rx}, our method and the concept of lifetime \textbf{are not restricted to moving antennas}, and can be applied to \textbf{any number of dynamic objects}, like a moving reflector or an antenna performing beamforming, but specific implementation details will not be covered in this paper.

\subsection{Multipath Structure}

While the \gls{em} fields are continuous everywhere in 3D, this continuity does not always hold in \gls{rt} simulations due to the nature of how ray paths are traced. Certain paths, such as the \gls{los} path, only exist within specific regions, and unless some smoothing technique is applied \cite{smoothing}, the collection of simulated paths tiles the space into continuous regions, separated by discontinuities.

Using the formalism defined in our previous work \cite{mpt}, we identify each ray path by a unique, ordered list of interacting objects, a list we call the ``path candidate''. Each element of this list specifies, in order, the type of interaction and the object involved, such as a specular reflection off a particular wall. The length of the list corresponds to the number of interactions the path undergoes between \gls{tx} and \gls{rx}. We define a \textbf{multipath cell} as the set of continuous regions where the set of valid path candidates---the multipath structure---remains unchanged (see \autoref{fig:multipath_cells}).

\subsection{How to identify and color cells}

The identification and coloring of multipath cells present certain implementation challenges, for which we provide the following solution.

Let the size of the scene, i.e., the number of objects (usually, triangular or rectangular facets), be \(N\), and the number of admissible interactions be \(K\). Then exhaustive \gls{rt}, which is what our tool performs by default, will process up to \(M = N(N-1)^{K-1}\) path candidates. Hence, each multipath cell, \(C_i\), can be identified by a vector, \(v_i \in \{0,1\}^M\), of boolean  values, with each entry indicating whether a path candidate is valid in that cell.

To reduce the memory footprint of our tool, we process the possible path candidates in smaller chunks, compressing the partial vectors into unique scalar values that may vary between snapshots. To ensure consistent coloring of multipath cells—which is necessary only for the visual output, but not for the metrics—we compress each multipath cell vector into a fixed-size space with the \gls {sha256}. If no hash collisions occur, as observed in our experiments, this enables us to map each cell to a constant unique color at a much lower memory cost.

\subsection{Related Metrics}

While a visual tool has definitively some advantages, especially for debugging purposes, it does not convey any concrete value that one can later use to make a decision.

For a given cell \(C_i\), we introduce the following metrics:

\begin{itemize}
  \item the area covered by each multipath cell, \(S_{i} = \text{area}(C_i)\);
  \item and the average minimal inter-cell distance, \(\overline{d_{i}}\);
\end{itemize}
where \(i\) indicates the index of the multipath cell, \(\overline{\cdot}\) is the ensemble average over a given cell, and the minimal inter-cell distance of an object \(x \in C_i\) is:

\begin{equation}
    d_i(x) = \min\limits_{y \notin C_i} \text{dist}(x, y),
\end{equation}
that is, the minimum distance the object \(x\) has to travel to leave the cell \(C_i\).

The area per multipath cell (in \si{\meter\squared}) provides insight into the size of the cells. However, as the same cell can be split among multiple spatial regions, as seen in \autoref{fig:multipath_cells}, the average minimal inter-cell distance is also important; it provides an estimate of the minimal lifetime of a multipath structure before an object transitions to another cell. Furthermore, a transition from one cell to another implies exactly one different path in the multipath structure, allowing for the reuse of all other paths, as they remain valid. Nevertheless, it may be computationally challenging to determine which path changes. In other words, the Hamming distance between two neighboring cell vectors, \(v_i\) and \(v_j\), is always 1.

\subsection{Urban Street Canyon}

\begin{figure}
  \centering
  \includegraphics[width=.6\columnwidth]{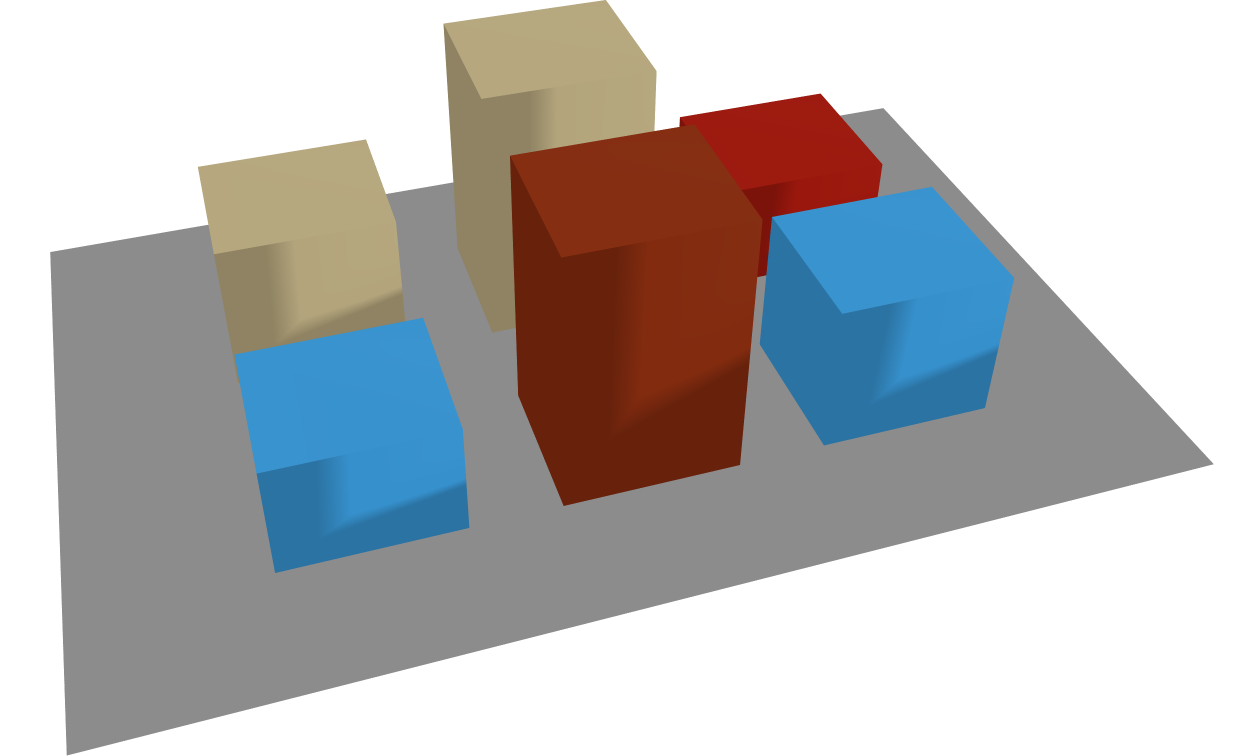}
  \caption{Street canyon scene from Sionna \cite{sionnart}.}
  \label{fig:street_canyon}
\end{figure}

For this example, we used the ``\emph{simple street canyon}'' scene from Sionna \cite{sionnart}, performing the simulation with our own open-source \gls{diffrt} toolbox, named DiffeRT\footnote{Repository: \url{https://github.com/jeertmans/DiffeRT}.}. A comprehensive tutorial\footnote{Tutorial: \url{https://differt.rtfd.io/eucap2025/notebooks/multipath.html}.} is also available to guide readers through the simulation procedure. This scene is spread over a \SI{120}{\meter} by \SI{185}{\meter} rectangular area, with building heights ranging from \qtyrange{20}{50}{\meter}. Each building in the scene has a square footprint of \SI{30}{\meter} by \SI{30}{\meter} and consist of triangular facets; we first merged these facets into quadrilaterals to ensure logical coherence in ray paths and align with the principles of \gls{dynrt}. This adjustment prevents path candidates from varying across individual triangular facets of the same building face. Furthermore, this \gls{sixb} scenario includes narrow streets (\SI{15}{\meter} wide) that intersect the main street canyon (\SI{20}{\meter} wide) at perpendicular angles. These narrow openings are frequently omitted, simplifying the model to represent the canyon as a long \gls{twob} scenario, where all building faces along each side of the street are treated as co-planar. Accordingly, we conducted an additional simulation with a simplified \gls{twob} scene, grouping buildings on each side of the main street canyon into single structures.

\paragraph{Simulation Parameters}

\begin{figure*}
  \centering
  \subfloat[Initial position.]{
    \begin{tikzpicture}[inner sep = 0pt, outer sep = 0pt]
      \node (map) {\includegraphics[width=0.3\textwidth]{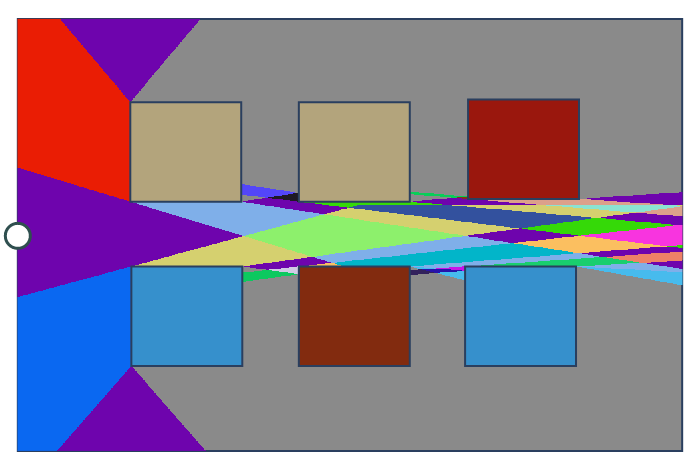}};
      \path (map.north) -- (map.south) node[pos=0.13, draw=white, very thick, fill=grey_color, inner sep=2pt, rounded corners, text=white] (los) {\footnotesize{\gls{los}}};
      \draw[<-, white, very thick, dashed] ([xshift=.6cm]map.west) to[out=90, in=180, looseness=1.5] (los);
    \end{tikzpicture}\label{fig:mlm_0}
  }
  \quad
  \subfloat[Intermediate position.]{\includegraphics[width=0.3\textwidth]{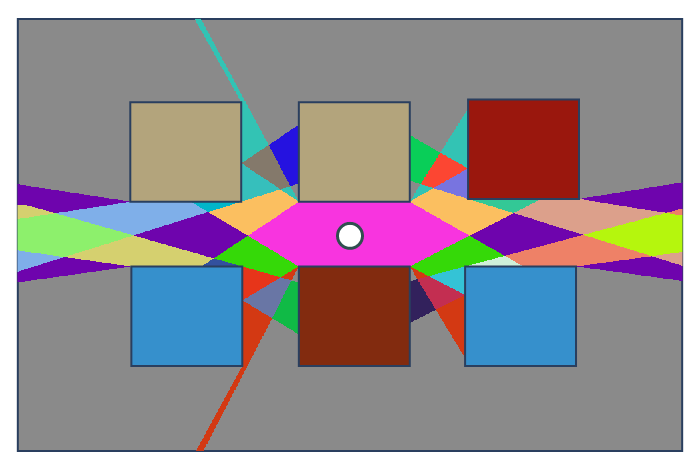}\label{fig:mlm_1}}
  \quad
  \subfloat[Final position.]{\includegraphics[width=0.3\textwidth]{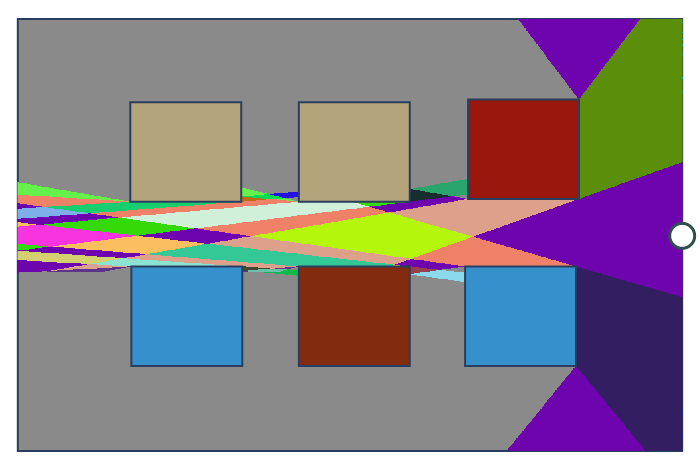}\label{fig:mlm_2}}
  \caption{\Glspl{mlm} for fixed \gls{tx} positions and a moving \gls{rx}. Each unique \(\text{MC}_i\) is assigned a unique color. E.g., the purple color indicates \gls{los}-only visibility. The same color is reused in other snapshots to indicate the same multipath cell. The no-multipath cell is transparent, thus leaving the ground apparent.}
  \label{fig:mlm_results}
\end{figure*}

\Gls{mlm} and metrics were derived from 50 \gls{rt} snapshots, each with \gls{tx} positioned at linearly spaced intervals between the scene's endpoints (see \autoref{fig:mlm_results}). \Gls{tx} was placed at a constant altitude of \SI{32}{\meter}, above all buildings, with a grid of 500-by-500 \Glspl{rx} uniformly distributed across the scene at an altitude of \SI{1.5}{\meter}. Only \gls{los} and first-order reflection paths were included in the simulation.

\paragraph{Results}

\begin{figure}
  \centering
  \begin{tikzpicture}[trim axis left, trim axis right]
    \begin{groupplot}[
      group style={
          group name=my plots,
          group size=1 by 2,
          vertical sep = 0pt,
      },
      ybar,
      width=0.9\columnwidth,
      height=4.0cm,
      xtick pos=left,
      ytick pos=left,
    ]
    \nextgroupplot[
      xlabel = {Cell area (\si{\meter\squared})},
      xticklabel pos=right,
      xlabel near ticks,
      xlabel style={inner sep=0pt},
      axis x line*=top,
      every y tick scale label/.style={
        at={(yticklabel* cs:0.95,0cm)},
        anchor=near yticklabel
      },
    ]
    \addplot +[
      hist={density, bins=30},
      draw=grey_color,
      fill=rose_color,
    ] table [y index=0] {data/area.txt};
    \addlegendentry{6-Building Street Canyon};
    \addplot +[
      hist={density, bins=30},
      draw=grey_color,
      fill=blue_color,
      opacity=0.5,
    ] table [y index=0] {data/area_simplified.txt};
    \addlegendentry{2-Building Street Canyon};
    \nextgroupplot[
      xlabel = {Average minimal inter-cell distance (\si{\meter})},
      xlabel near ticks,
      xlabel style={inner sep=0pt},
    ]
    \addplot +[
      hist={density, bins=30},
      draw=grey_color,
      fill=rose_color,
    ] table [y index=0] {data/dist.txt};
    \addplot +[
      hist={density, bins=30},
      draw=grey_color,
      fill=blue_color,
      opacity=0.5,
    ] table [y index=0] {data/dist_simplified.txt};
  \end{groupplot}
  \path (my plots c1r1.south west) -- ++(-1, 0) node[rotate=90] {Density (mass normalized)};
  \end{tikzpicture}
  \caption{Histograms of (top) area occupied by each multipath cell and (bottom) average distance a \gls{rx} has to travel to leave its current cell.}
  \label{fig:histograms}
\end{figure}
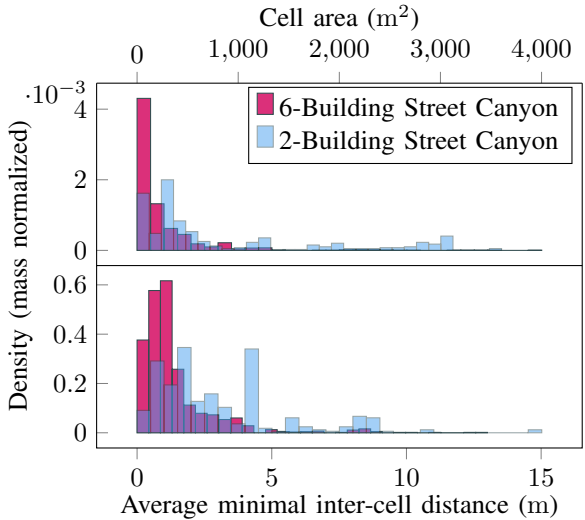

\autoref{fig:mlm_results} illustrates \gls{mlm} at three different \gls{rx} positions. Using data from the 50 snapshots, we compiled a histogram of \(S_{i}\) and \(\overline{d_{i}}\) shown in \autoref{fig:histograms}. \autoref{tab:results} summarizes the mean and median values of all measured \(S_{i}\) and \(\overline{d_{i}}\).

\begin{table}
  \renewcommand{\arraystretch}{1.3}
  \caption{Mean and median values over all 50 simulation snapshots and cells.}
  \label{tab:results}
  \centering
  \begin{tabular}{l|c|c|c|c}
    & \multicolumn{2}{c|}{\gls{sixb}} & \multicolumn{2}{c}{\gls{twob}} \\
    \hline
    & \(S\) (\si{\meter\squared}) & \(\overline{d}\) (\si{\meter}) & \(S\) (\si{\meter\squared}) & \(\overline{d}\) (\si{\meter}) \\
    \hline
    Mean  & 225.62 & 1.40 & 840.56 & 3.08 \\
    \hline
    Median & 86.43 & 1.00 & 371.38 & 2.30  \\
  \end{tabular}
\end{table}

\paragraph{Comments}

Even in this simple scenario with limited paths, \autoref{fig:mlm_results} already exhibits the complex spatial distribution of multipath cells. Increasing the reflection order would only introduce more cells and further reduce the mean values of \(S_i\) and \(\overline{d_{i}}\). However, diffraction would not affect these cells, as a diffraction path of any order exists wherever a corresponding reflection-only path is present. Assuming disk-shaped cells, the \gls{sixb} scenario yields a cell radius from \qtyrange{5.25}{8.47}{\meter} (median to mean), aligning with spatial consistency values of \qtyrange{5}{10}{\meter} in local areas \cite{consistency}. However, cells are rarely disk-shaped, as shown by the relatively small \(\overline{d}\) and observed in \autoref{fig:mlm_results}; they tend to extend along a primary direction with the longest lifetime. Ultimately, as demonstrated by the \gls{twob} scenario, simplifying the scene can extend the lifetime of a given multipath structure, further enhancing \gls{dynrt}'s suitability.

\section{Conclusion}

In this work, we have presented a comparative analysis of \gls{dynrt} and \gls{diffrt} methods for \gls{rt}, using 3DSCAT and Sionna as respective examples. While \gls{dynrt} offers manual differentiation, providing interpretability through symbolic derivatives, it is constrained by the tedious derivation process and limited variable applicability. Conversely, \gls{diffrt} leverages \gls{ad} to calculate derivatives for nearly any scene parameter with ease and flexibility, though often at the cost of restricted operational primitives and interpretability.

\balance

Despite the unique benefits of both approaches, the availability of tools implementing \gls{dynrt} or \gls{diffrt} remains limited. The development of our open-source framework, DiffeRT, aims at offering a \gls{rt}-focused alternative to Sionna. Ultimately, expanding the range of open-source \gls{dynrt} and \gls{diffrt} tools will be crucial for advancing the field of \gls{rt}.

Then, we have introduced a novel simulation-based method, via \gls{mlm} and two metrics, to investigate the scope of applications of extrapolating \gls{rt} snapshots, as used in \gls{dynrt}. The method shows that scene simplifications can increase the scope where \gls{dynrt} is best suited, but also highlight the relative complex shapes of multipath cells even in simple scenarios. In future research, we intend to expand the scope of the study to encompass larger scenes and additional categories of dynamic objects, including multiple moving reflectors.

\bibliographystyle{IEEEtran}
\bibliography{IEEEabrv,biblio}

\end{document}

%% file: tikz/pipeline.tex
\begin{tikzpicture}  %
    \node[block] (rt) at (0,0) {Tracing of\\ray paths};
    \node[block] (em) at (2.5,0) {\gls{em} fields\\computation};
    \draw[<-,thick, rounded corners] (rt.west) -| ++(-.5, -.75) node[below, align=center] {Input scene\\(objects, \gls{tx}, \gls{rx}, ...)};
    \draw[->, thick] (rt.east) -- (em.west);
    \draw[->, thick] (rt.east) -- (em.west) node[midway] (midway) {};
    \draw[->, thick] (midway.center) -- ++(0, -.75) node[below, align=center] {All rays from\\\gls{tx} to \gls{rx}};
    \draw[->, thick, rounded corners] (em.east) -- ++(+.5, 0) -- ++(0, -.75) node[below, align=center] {\gls{em} fields at\\\gls{rx}};
\end{tikzpicture}

%% file: tikz/multipath_cells.tex
\DeclareRobustCommand{\show_multipath_cells}[1]{
\begin{tikzpicture}
    \coordinate (SW) at (-0.5, -1.5);
    \coordinate (NE) at (+2.5, +1.5);
    \coordinate (S1) at (0.5, +0.75);
    \coordinate (E1) at (1.5, +0.75);
    \coordinate (S2) at (0.25, -0.75);
    \coordinate (E2) at (1.25, -0.75);
    \coordinate (TX) at (0, 0);

    \clip[rounded corners] (SW) rectangle (NE);

    \ifnum1=\ifnum#1=0 1\else#1\fi
    \begin{scope}
        \coordinate (IS1) at (1.0, +1.5);
        \coordinate (IE1) at (4.0, +1.5);
        \coordinate (IS2) at (0.5, -1.5);
        \coordinate (IE2) at (2.5, -1.5);
        \clip (SW) --  (SW |- NE) -- (IS1) -- (S1) -- (E1) -- (IE1) -- (NE) -- (IE2) -- (E2) -- (S2) -- (IS2) -- cycle;
        \fill[los_color,fill opacity=.5] (SW) rectangle (NE);
    \end{scope}
    \fi

    \ifnum2=\ifnum#1=0 2\else#1\fi
    \begin{scope}
        \coordinate (IS1) at (2.0, -1.5);
        \coordinate (IE1) at (4.0, +0.0);
        \clip (S1) --  (IS1) -- (SW -| NE) -- (IE1) -- (E1) -- cycle;
        \fill[rup_color,fill opacity=.5] (SW) rectangle (NE);
    \end{scope}
    \fi

    \ifnum3=\ifnum#1=0 3\else#1\fi
    \begin{scope}
        \coordinate (IS2) at (0.75, +0.75);
        \coordinate (IS2bis) at (2., +1.5);
        \coordinate (IE2) at (2.5, +0.0);
        \clip (S2) --  (IS2) -- (E1) -- (IS2bis) -| (IE2) -- (E2) -- cycle;
        \fill[rdw_color,fill opacity=.5] (SW) rectangle (NE);
    \end{scope}
    \fi

    \draw[dashed, ultra thick, rounded corners] (SW) rectangle (NE);
    \draw[ultra thick] (S1) -- (E1);
    \draw[ultra thick] (S2) -- (E2);
    \node[draw=tx_color, circle, ultra thick, fill=white] (TX) {};

\end{tikzpicture}
}